\documentclass[a4paper,11pt]{article}
\pdfoutput=1 

\usepackage{jheppub} 

\usepackage{amssymb}
\usepackage{amsmath}
\usepackage{tikz}
\usepackage{graphicx}
\definecolor{uibred}{RGB}{167, 38, 47}

\def\del{\partial}

\newcommand{\eqn}[1]{Eq.~\eqref{#1}}

\long\def\comment#1{ }

\def\0{{\boldsymbol 0}}

\def\xs{x_\text{br}}
\def\coh{\text{coh}}

\def\Pc{{\cal P}}
\def\Kc{{\cal K}}
\def\Oc{{\cal O}}

\newcommand{\beq}{\begin{eqnarray}}
\newcommand{\eeq}{\end{eqnarray}}

\newcommand{\be}{\begin{eqnarray*}}
\newcommand{\ee}{\end{eqnarray*}}

\newcommand{\rmd}{{\rm d}}

\newcommand{\rmi}{i}

\def\abar{{\rm \bar\alpha}}

\def\br{{\rm br}}

\def\glu{{\rm g}}
\def\glug{{\rm gg}}
\def\gq{{\rm gq}}
\def\qg{{\rm qg}}
\def\qq{{\rm qq}}
\def\NS{{\rm NS}}
\def\S{{\rm S}}
\def\q{{\rm q}}
\def\qbar{{ \rm \bar q}}
\newcommand{\nn}{\nonumber\\ }

\newcommand{\qhat}{\hat{q}}

\begin{document}

\title{Universal quark to gluon ratio in medium-induced parton cascade}

\author[a,b]{Yacine Mehtar-Tani}
\affiliation[a]{Institute for Nuclear Theory, University of Washington, Box 351550,\\ Seattle, WA 98195-1550, USA}
\affiliation[b]{Physics Department, Brookhaven National Laboratory, Upton, NY 11973, USA}

\author[c]{Soeren Schlichting}
\affiliation[c]{Department of Physics, University of Washington, Box 351560,\\ Seattle, WA 98195-1560, USA}



\abstract{We investigate the radiative break-up of a highly energetic quark or gluon in a high-temperature QCD plasma. Within an inertial range of momenta $T  \ll \omega \ll E$,  where $E$ denotes the energy of the original hard parton (jet) and $T$ the temperature of of the medium, we find that, as a result of the turbulent nature of the underlying parton cascade, the quark to gluon ratio of the soft fragments tends to a universal constant value that is independent of the initial conditions. We discuss implications of this result to jet quenching physics and the problem of thermalization of the quark-gluon plasma in heavy ion collisions.}

\date{\today}
\maketitle
\flushbottom


\section{Introduction}
\label{sec:Introduction}

In the early 80's, Bjorken suggested that the suppression of  jets would signal the formation of a Quark-Gluon Plasma (QGP) in hadronic collisions~\cite{Bjorken:1982}. Two decades later, the RHIC experiment successfully observed jet quenching on high-$p_T$ hadron spectra \cite{Adcox:2001jp,Adler:2002xw}, initiating a program that was carried over at the LHC \cite{Aad:2010bu,Chatrchyan:2012nia,Aad:2014bxa,Abelev:2013kqa}, using reconstructed jets to investigate the deconfined matter created in these collisions by studying medium modifications of jet observables in nucleus-nucleus collisions as compared to proton-proton collisions. 

Understanding how a jet evolves as a multi-partonic system is crucial to quantitatively probe the QGP.  However, although jet evolution in the vacuum is well established theoretically \cite{Ali:2010tw,Sapeta:2015gee}, the theory of jet energy loss in the presence of a dense medium remains an active field of study (see \cite{Mehtar-Tani:2013pia,Blaizot:2015lma} for recent reviews). Because of its multi-scale nature, spanning a wide range of scales from $\sim 1$ GeV to  $ \sim 1 $ TeV, jet quenching physics is also instrumental in understanding dissipation of energy in a colored medium and hence, non-equilibrium QCD dynamics underlying the formation of the QGP in hadronic collisions.  
In effect, the relatively hard particles, produced in such collisions, experience multiple interactions in the aftermath of the collision causing their radiative break-up into softer fragments, which eventually thermalize to form an equilibrated QGP \cite{Baier:2000sb,Kurkela:2018vqr}. 



Previous theoretical studies, in both problems of jet evolution and thermalization in heavy ion collision, focused mostly on the pure gluonic dynamics, neglecting quark degrees of freedom for simplicity. This is further motivated by the fact that the leading order QCD splitting function is dominated by gluons in the soft sector where multiple parton branchings are expect to be relevant. However, it was recently pointed out that the medium-induced cascade is characterized by democratic branchings \cite{Blaizot:2013hx,Blaizot:2015jea}, therefore, one should expect quarks to play a significant role in the energy transport from highly energetic partons in the jets to the medium temperature scale. The purpose of this work is to investigate this question and understand how quarks and gluons couple through the medium-induced cascade. 

Our discussion is organized as follows: We introduce the coupled evolution equations for in-medium jet fragmentation in Sec.~\ref{sec:Analytics}, and discuss the properties of analytic solutions at early times and in the stationary turbulent regime. We present numerical results for the in-medium fragmentation functions of quark and gluon jets in Sec.~\ref{sec:Results} and further discuss the chemical composition of fragments and the energy loss of the jet. We conclude in Sec.~\ref{sec:Conclusions}, with a brief summary of our findings, along with a discussion of possible phenomenological consequences of our work and implications for the study of quark production during the pre-equilibrium stage of high-energy collisions. Details on 
stationary turbulent solutions for forced cascades are provided in \ref{sec:AppendixStationaryTurbulence}.


\section{Medium-induced quark-gluon cascade and wave turbulence}
\label{sec:Analytics}

A QCD jet that emerges from a hard collision evolves from a highly energetic parton to a more complex object via multiple collinear branchings. Such a parton cascade is characterized by a strong decrease of the virtuality of the intermediate partonic states, which due to the collinear nature of the process are highly collimated. In vacuum this evolution is described by DGLAP type evolution equations \cite{Altarelli:1977zs} and proceeds down to non-perturbative scales where confining forces take over. 


However, in the presence of a hot QCD medium another cascading process is triggered by the interaction of the jet with the colored constituents of the medium. Multiple interactions with the medium can induce the emission of Bremsstrahlung radiation, which is characterized by a coherence length $t_\coh$ that corresponds to the time it takes for the emitted quanta to form via a diffusion process in transverse momentum space (with respect to the direction of the jet propagation). Based on the uncertainty principle, a gluon emitted with frequency $\omega$ and transverse momentum $k_\perp$ forms at $t \sim \omega/k_\perp^2$. During this time it accumulated $k_\perp^2 \sim \qhat \, t $ transverse momentum square, where $\hat q \equiv \rmd \langle k_\perp^2\rangle/\rmd t \sim m_D^2/\ell_\text{mfp}$, the so-called jet-quenching parameter, is the relevant diffusion coefficient, with $\ell_\text{mfp}\sim (g^2 T)^{-1}$ and $ m_D^2\sim g^2 T^2$, the in-medium mean-free-path and the Debye screening mass in a thermal plasma, respectively. Solving the latter equations self-consistently one finds, $t_\coh = \sqrt{\omega/\hat q }$ for the coherence time.  

For short formation times $t_\coh \ll L$, the medium-induced radiation spectrum scales parametrically as \cite{Baier:1996kr,Baier:1996sk,Baier:1998kq,Zakharov:1996fv},
\beq
\label{eq:dIdw}
\left.\omega  \frac{\rmd I}{\rmd \omega}\right|_{\rm Medium} \sim \alpha_s C_R \frac{L}{t_\coh(\omega)} \sim \sqrt{ \frac{\omega_s }{\omega}},
\eeq 
where $L$ denotes the length of the medium, $\omega_s = \alpha_s^2 \hat q L^2$, the characteristic scale for multiple branchings and $C_R$ is the corresponding color factor, i.e. $C_R=C_F =(N_c^2-1)/2N_c$ and $C_R=C_A=N_c$ for radiation off a quark and  a gluon, respectively. Since the medium-induced gluon accumulates a finite transverse momentum $k_\coh \sim (\omega \hat q)^{1/4}$ over the course of the emission process, the collinear (mass) singularity is regulated for medium induced radiation. However, in contrast to vacuum radiation which is always tied with the hard collision vertex, the emission rates for medium induced radiation are enhanced by the length of the medium $\frac{L}{t_\coh(\omega)}$ as emissions can occur anywhere throughout the medium with equal probability (for a homogeneous medium). In this case, the traversed path-length $L$ naturally plays the role of an ordering variable and multiple emissions with $t_\coh \ll L$ can be ``resummed" via an effective kinetic description.

We emphasize that, unlike the DGLAP splitting kernels \cite{Altarelli:1977zs}, that are scale invariant (up to logarithmic factors) the in-medium emission rate \eqn{eq:dIdw} decreases as a function of the parton energy as $1/\sqrt{\omega}$. This is an essential feature of the above elementary process that will qualitatively impact the properties of the in-medium parton cascade. The characteristic decrease of the spectrum as a function of the parton energy, follows directly from 
the increasing formation time and can be understood as a consequence of the QCD analog of the Landau-Pomerantchuk-Migdal (LPM) effect \cite{Baier:1996kr,Baier:1996sk,Baier:1998kq,Zakharov:1996fv,Wiedemann:2000za}: During the radiation process the system interacts with many scattering centers that act coherently as a single one. Hence, the effective number of scatterers decreases at larger $\omega$ up to the the characteristic frequency $\omega_c \equiv \hat q L^2$ where  $t_\coh \sim L$. Conversely, for $\omega>\omega_c$ where $t_\coh \, > L$, medium induced radiation is strongly suppressed as the jet cannot  ``resolve" the medium from the hard vertex.

We also note that due to the absence of a collinear singularity, soft gluons, that are copiously produced via medium-induced radiation, tend to be radiated at large angles $\theta_\coh \sim (\hat q/\omega^3)^{1/4} $. This effect is further enhanced by the cascading process, as a result, the medium-induced cascade develops at larger angles compared to vacuum like radiation \cite{Blaizot:2014ula,Kurkela:2014tla,Blaizot:2014rla}.

\subsection{Coupled evolution equations for in-medium jet fragmentation}

We will, from now on, focus exclusively on medium induced radiation, and consider a highly energetic and on-shell parton, of energy $E \gg T$, propagating through a hot QCD medium with temperature $T$.  Its interaction with the medium constituents, as alluded to above,  causes the initial parton to successively branch into an arbitrary number of partons  transporting its energy from high to low frequencies.  In the infinite medium limit this branching process is Markovian, that is, the probability for a single parton to branch into $N$ partons reduces to $N-1$ independent and quasi-instantaneous elementary $1 \to 2 $ splitting processes.  Generalizing  \eqn{eq:dIdw} beyond the soft limit, the corresponding splitting rate takes the generic form 
\beq\label{eq:br-rate}
\frac{\rmd \Gamma}{\rmd z } \equiv \frac{\rmd \Pc}{\rmd z\rmd t } = \frac{ \Kc(z)}{t_\br(\omega)} \,, \quad\text{with} \quad t_\br(\omega) \equiv  \frac{1}{\abar}  \sqrt{\frac{\omega}{\hat{\bar{q}}}},
\eeq
where $\omega < E$ is the energy of the parton, inside the cascade, that splits into two daughters carrying the fractions $z$ and $1-z$ of its energy and $\bar \alpha\equiv \alpha_s/\pi$.  Note that for future convenience, we have stripped $\hat q $ from its color factor by introducing the reduced jet-quenching parameter $\hat {\bar q} \equiv \hat q / C_R $, and absorbed all color factors into the splitting kernel $\Kc(z)$ (cf. \eqn{eq:split-def}).

 \eqn{eq:br-rate} is valid in the LPM regime, that is, when $\ell_\text{mfp} \ll t_\coh(\omega) \ll L$. The lower limit of the coherence length, $t_\coh(\omega) \sim\ell_\text{mfp}$, corresponds to the single scattering regime and is known as the Bethe-Heitler regime that is characterized by the frequency $\omega_\text{BH} \equiv \hat q \,\ell_\text{mfp}^2 \sim T$.  At this scale, $\omega \sim T$, inverse parton merging processes, as well as elastic processes also become important \cite{Jeon:2003gi,Ghiglieri:2015ala} and will be responsible for equilibrating the influx of energy and particles from the hard sector~\cite{Baier:2000sb,Kurkela:2014tea,Kurkela:2018vqr}. However, so long as the energies of the particles inside the cascade are much larger than the temperature of the plasma ($\omega \gg T$) these contributions are power suppressed and will therefore be neglected in our analysis. 

The observable that we shall investigate in order to characterize the medium-induced cascade is the inclusive in-medium parton distribution (or fragmentation function),
\beq
D_\rmi(x,\tau) \equiv x \frac{\rmd N_\rmi}{\rmd x},
\eeq
where $x=\omega/E$ denotes the energy fraction carried by a parton of frequency $\omega$ w.r.t to the initial energy $E$ of the original parton/jet. Here, $\rmi = g,q$ or $ \bar q$ labels the species of the measured parton and we have introduced the dimensionless time variable
\beq 
\tau \equiv \frac{t}{t_\br(E)} = \abar \sqrt{\frac{\hat{\bar{q}}}{E}} t,
\eeq 
which fully accounts for the jet energy $(E)$ dependence of the radiative break-up process.

For the quark (anti-quark) components we shall adopt the standard  flavor singlet (S) and non-singlet (NS) decomposition: 
\beq
 D_\S \equiv \, \sum_{i=1}^{N_f} \, (D_{\q_i} +D_{\qbar_i}) \quad \text{and} \quad D_\NS^{(i)} \equiv \, D_{\q_{i}} - D_{\qbar_{i}},
\eeq
where $N_f$ is the number of active massless quark flavors, which together with $D_\glu$ obey the following set of coupled evolution equations:

\beq\label{eq:evol-eq-glu}
\frac{\del }{\del \tau } D_\glu\left(x,\tau\right)&=&\int_{0}^{1} dz \, \Kc_\glug(z) \left[ \sqrt{\frac{z}{x}} D_\glu\left(\frac{x}{z} \right)- \frac{z}{\sqrt{x}} D_\glu(x) \right] -   \int_{0}^{1} \rmd z \,K_\qg(z)  \frac{z}{\sqrt{x}}\, D_\glu\left(x\right)\nn
&+&\int_0^1 \rmd z K_\gq(z)  \,  \sqrt{\frac{z}{x}} \,D_\S \left(\frac{x}{z}\right),\nn
\eeq
\beq\label{eq:evol-eq-S}
\frac{\del }{\del \tau } D_\S\left(x,\tau\right)&=& \int_{0}^{1} dz \, \Kc_\qq(z) \left[ \sqrt{\frac{z}{x}} D_\S\left(\frac{x}{z} \right)- \frac{1}{\sqrt{x}} D_\S(x) \right] + \int_{0}^{1} dz \, \Kc_\qg(z) \, \sqrt{\frac{z}{x}} D_\glu\left(\frac{x}{z}\right)\nn
\eeq
and
\beq\label{eq:evol-eq-NS}
\frac{\del }{\del \tau } D_\NS^{(i)}\left(x,\tau\right)&=& \int_{0}^{1} dz \, \Kc_\qq(z) \left[ \sqrt{\frac{z}{x}} D_\NS^{(i)}\left(\frac{x}{z} \right)- \frac{1}{\sqrt{x}} D_\NS^{(i)}(x) \right]
\eeq
where to leading logarithmic accuracy (in $\log(E/T)$) the various splitting kernels are given by \cite{Arnold:2008zu,Ghiglieri:2015ala}
\begin{eqnarray}\label{eq:split-def}
\Kc_\glug(z)  &=& \frac{1}{2}~2C_{A} \frac{[1-z(1-z)]^2}{z(1-z)}~\sqrt{\frac{(1-z)C_A+z^2C_A}{z(1-z)}}\;,\\
\Kc_\qg(z) &=& \frac{1}{2}~2 N_f T_{F} \Big(z^2+(1-z)^2\Big)~\sqrt{\frac{C_F-z(1-z)C_{A}}{z(1-z)}}\;, \\
\Kc_\gq(z) &=& \frac{1}{2}~C_{F} \frac{1+(1-z)^2}{z}~\sqrt{\frac{(1-z)C_A+z^2C_F}{z(1-z)}}\;, \\
\Kc_\qq(z)  &=& \frac{1}{2}~C_{F} ~\frac{1+z^2}{(1-z)}~\sqrt{\frac{zC_A+(1-z)^2C_F}{z(1-z)}}\;. 
\end{eqnarray}

The collision integrals in the r.h.s of the above set of equations comprise (positive) gain and (negative) loss terms that correspond, respectively,  to the production of a parton $x$ from the splitting of a parent parton with energy fraction $x/z$, and the decay of a parton $x$ into softer fragments $zx$ and $(1-z)x$. This construction ensures the conservation of the total jet energy i.e., 
\beq
\label{eq:EnergyConsEq}
\epsilon(\tau)= \int_0^1 \rmd x  \, \left ( D_\glu(x,\tau)+D_\S(x,\tau) \right) =1. 
\eeq
in the absence of a finite flux at the $x\to0$ boundary. However, as we shall see shortly, a non-vanishing flux of energy builds up immediately due the fact that the rate of successive branchings (\ref{eq:br-rate}) increases along the cascade transporting energy from hard $x\sim1$ scales down to arbitrarily soft scales $x\to0$.  Indeed it has been shown analytically~\cite{Blaizot:2013hx} that for a simplified version of the kinetic equations -- considering only gluons along with a simplified form of the splitting Kernel -- the energy decreases as $\epsilon(\tau)=e^{-\pi\tau^2}$ during the radiative break-up cascade. Physically, energy flows without accumulating down to the soft scale $x\sim T/E \ll 1$ where it is dissipated in the thermal medium \cite{Baier:2000sb}. 

\subsection{Single emission spectra \& Breakdown of small $\tau$ expansion}
Before we analyze the full dynamics of the radiative break-up process, it is instructive to investigate the above equations perturbatively, that is, to compute the $\Oc (\tau)$ correction to the initial condition
\beq\label{eq:initial-cond-g}
D_\glu(x,\tau=0) =\delta(1-x),   \quad D_\S(x,\tau=0) =D_\NS(x)= 0,
\eeq
that describes a single gluon that initially carries all of the jet energy.  Since the non-singlet distribution does not receive contributions from the other channels it vanishes identically to all orders, i.e., $D_\NS(x)=0$. By inserting, \eqn{eq:initial-cond-g} into Eqs.~(\ref{eq:evol-eq-glu}) - (\ref{eq:evol-eq-NS}), we readily obtain 
\begin{eqnarray}\label{eq:lo-glu}
&& D_\glu (x,\tau)\, \simeq \,\delta(1-x)+ \left [ x \Kc_\glug(x)   - \int_0^1 \rmd z z \left(\Kc_\glug(z) +\Kc_\qg(z)\right)  \, \delta(1-x) \right] \tau   \nn
 &&   D_\S (x,\tau)\, \simeq \, x K_\qg(x)\, \tau  \;,
\end{eqnarray}
such that at low momentum, one finds
\begin{eqnarray}\label{eq:lo-glu-smallx}
D_\glu (x,\tau)\simeq \frac{C_A^{3/2} \,  \tau }{ \sqrt{x}}, \quad \text{and} \quad   D_\S (x,\tau)\simeq \frac{1}{2} \, 2N_f \, T_R\, C_F^{1/2}\,  \, \tau  \;\sqrt{x}.
\end{eqnarray}
Similarly, one obtains the perturbative solution for the the distributions of partons inside a quark jet,
\begin{eqnarray}
\label{eq:lo-qua}
&& D_\NS (x,\tau)=D_\S (x,\tau)= \,  \simeq \,\delta(1-x)+ \left [ x \Kc_\qq(x)   - \int_0^1 \rmd z z \Kc_\qq(z) \, \delta(1-x) \right]\,\tau\nn
&&   D_\glu(x,\tau)\, \simeq \, x \Kc_\gq(x) \, \tau  \;. 
\end{eqnarray} 
where we have used the following initial condition 
\beq\label{eq:initial-cond-q}
D_\glu(x,\tau=0) =0,   \quad D_\S(x,\tau=0) =D_\NS(x)= \delta(1-x),
\eeq
that describes a single quark that initially carries all of the jet energy. Energy conservation as in Eq.~(\ref{eq:EnergyConsEq}) is satisfied at this order for both quark and gluons jets -- however as we will discuss later in Sec.~\ref{sec:Results} a finite energy flux is generated already at the next order $\Oc (\tau^2)$.

Even though the perturbative solutions in Eqs.~(\ref{eq:lo-glu}) and (\ref{eq:lo-qua}) are formally accurate to $\Oc (\tau)$, their range of validity is also limited in $x$ space. Considering for instance the gluon component in Eq.~(\ref{eq:lo-glu}),  the  $\Oc (\tau)$ correction to the distribution diverges as  $D_\glu (x,\tau) \sim (1-x)^{- 3/2}$, indicating that a non-perturbative treatment is required in the $x\simeq 1$ region. Of course, this is not surprising as analytic solutions of simplified models show the existence of an essential singularity when $x\to1$~\cite{Blaizot:2013hx}.

Similarly, in the small $x\simeq 0$ region, higher order corrections quickly become important due the fact that the rate for subsequent splittings, initiated by fragments with $x<1$ are enhanced by a factor of $1/\sqrt{x}$ relative to the splitting rate off the original hard parton ($x=1$).  We can further quantify this behavior, by analyzing the probability $P_{\rm split}$ for a quark/gluon with momentum fraction $x$ to undergo further splitting. Considering the $g\to q\bar{q}$ splitting we can simply compute
\begin{eqnarray}
\label{eq:PSplitQG}
\Pc^{\rm split}_\qg= \frac{\tau}{\sqrt{x}} \int_{0}^{1} \rmd z~\Kc_\qg(z) \stackrel{N_f=3}{\simeq} 3.54\frac{\tau}{\sqrt{x}}
\end{eqnarray}
which for any $\tau>0$ becomes of order one for sufficiently small values $x \lesssim  \tau^2$. Similar conclusions can be reached for additional gluon emissions of a quark/gluon, although in this case there is subtlety pertaining to the soft divergence of the integral of the kernel when $z\to 1$ or $z \to 0$. In order see that, it proves insightful to distinguish between quasi-democratic ($z\sim 1/2$) and very asymmetric splittings ($z\ll 1$ or $1-z \ll 1$). Separating for example the $\glu \to \glu \glu$ process into splittings where $\text{min}(z,1-z)>z_{\rm min}$, such that both fragments have momentum fraction larger than $z_{\rm min}$  
\begin{eqnarray}
\Pc^{\rm split}_{\glug}\Big(z>z_{\rm min}\Big) = \frac{\tau}{\sqrt{x}} \int_{z_{\rm min}}^{1-z_{\rm min}} \rmd z~\Kc_\glug(z) 
\end{eqnarray}
one finds that the probability for quasi-democratic splittings $(z_{\rm min} \sim 1/2)$, exhibits the same parametric dependence as in Eq.~(\ref{eq:PSplitQG}), i.e. 
\begin{eqnarray}\label{eq:PSplitGG-demo}
\left.\Pc^{\rm split}_{\glug}\Big(z>z_{\rm min}\Big) \right|_{z_{\rm min} \sim 1/2}  \simeq \Kc_\glug(1/2) (1-2 z_{\rm min}) \frac{\tau}{\sqrt{x}}\;,  
\end{eqnarray}
with $\Kc_\glug(1/2)=9 \sqrt{3} C_{A}^{3/2} / 4$. Despite the fact that the splitting probability for very soft splittings $(z_{\rm min} \ll 1)$ is divergent
\begin{eqnarray}
\left.\Pc^{\rm split}_{\glug}\Big(z>z_{\rm min}\Big) \right|_{z_{\rm min} \ll 1}  \simeq \frac{4C_{A}^{3/2}}{\sqrt{z_{\rm min}}} \frac{\tau}{\sqrt{x}}\;,
\end{eqnarray}
the effects of such very soft splittings on the distribution are small, as the energy of the emitting particle is not changed appreciably, which manifest itself in a cancellation of the soft divergence between the gain and loss terms. A more careful analysis reveals that the contributions to the evolution equations of the fragmentation function from the limit of $z_{\rm min} \ll1$, can be cast in the form of a diffusion equation
\begin{eqnarray}\label{eq:diff-rate}
\left.\frac{\del }{\del \tau } D_\glu\left(x,\tau\right) \right|_{z_{\rm min} \ll 1}&=&\int_{1-z_{\rm min}}^{1} \rmd z~\Kc_\glug(z) \left[ \sqrt{\frac{z}{x}} D_\glu\Big(\frac{x}{z}\Big) - \frac{z}{\sqrt{x}} D_\glu(x) \right]  - \int_{0}^{z_{\rm min}} \rmd z~\Kc_\glug(z) \frac{z}{\sqrt{x}} D_\glu(x)  \nonumber \\   
&=& \frac{C_A^{3/2}}{\sqrt{x}}~\sqrt{z_{\rm min}} \left[2x \frac{\partial}{\partial x}D_{\glu}(x) -  D_\glu(x) \right] + \Oc\Big(z_{\rm min}\Big)^{3/2}\;,
\end{eqnarray}
where the $1/\sqrt{z_{\rm min}}$ divergence has cancelled out between the gain and loss terms as anticipated. Instead the effective rate for soft radiations scales as, $ \sim \sqrt{z_{\rm min}/x}$, as one can read off the first term in the r.h.s. of the above equation. Because of its $\sqrt{z_{\rm min}} \ll 1$ scaling, the latter rate is small compared to the democratic branching rate $\propto (1-2 z_{\rm min})\sim 1$, (cf. \eqn{eq:PSplitGG-demo}), demonstrating that asymmetric branchings are sub-dominant. Note that this conclusion is closely tied to the power of the divergence of the splitting  kernel that turns out to be mild enough not to affect the qualitative features of the cascade that will be discussed in the next section. On the other hand, for a kernel that diverges like $1/z^2$ or faster, the first term in the r.h.s. of \eqn{eq:diff-rate} would be constant or divergent, respectively. In this case, strongly asymmetric branchings would constitute the dominant processes. 

Most importantly, one observes that all of the above splitting rates scale parametrically as $\propto 1/\sqrt{x}$ as a function of the momentum fraction $x$ of the emitter. Based on our analysis, one therefore concludes that there is a dynamically generated scale $\xs \sim \tau^2$, below which the single emission approximation breaks down. Hence, for $x<\xs$ the fragmentation function is  dominated by multiple successive emissions which can lead to dramatic changes in the spectra.

%
%
\subsection{Stationary Kolmogorov solution \& turbulent energy/particle flux}
\label{sec:KolmogorovSolution}

Below the scale $x_\br$, the energy transfer from the hard ($x\sim1$) partons inside the jet to the soft thermal medium ($x\sim T/E$) proceeds via multiple quasi-democractic branchings. Due to the characteristic energy dependence of the elementary branching process, the splitting rates for $x\lesssim x_\br$ become large, i.e. $\Gamma^{\rm split} \tau \gtrsim 1$, and one may therefore expect that the distribution of partons in the infrared approach a fixed point solution in order to prevent an unphysical rapid change of the distribution. Even though a stationary fixed point solution can not be achieved due to the varying influx of energy from the hard sector $(x\gg x_\br)$, such variations occur on time scales which are large compared to the local interaction rates and one may therefore still expect the local approach to a fixed point solution in the small $x$ region. Having this in mind, we shall investigate, as an intermediate step, the fixed points of the evolution equations Eqs.~(\ref{eq:evol-eq-glu}) - (\ref{eq:evol-eq-NS}) in order to gain further analytic insight into the structure of solution for $x\lesssim x_\br$.

Guided by the analytic treatment of purely gluonic models~\cite{Blaizot:2013hx,Blaizot:2015jea}, it is natural to search for stationary non-equilibrium solutions of the form 
\begin{eqnarray}\label{eq:qg-spect}
D_\glu(x)=\frac{G}{\sqrt{x}}\;, \qquad D_\S=\frac{Q}{\sqrt{x}}\;. 
\end{eqnarray}
which, as we will discuss shortly, correspond to the Kolmogorov-Zakharov (KZ) spectra associated with an (inverse) energy cascade. By inserting the above ansatz in Eqs.~(\ref{eq:evol-eq-glu}) - (\ref{eq:evol-eq-NS}) and requiring that  
\beq
\label{eq:statcon}
\partial_{\tau} D_\glu(x)=\partial_{\tau} D_\S(x)=0,
\eeq
we find that the chemistry of fragments is uniquely determined by the balance of the $g\to q\bar{q}$ and $q\to qg$ processes, which gives rise to an algebraic constraint
\begin{eqnarray}\label{eq:qg-ratio}
\frac{Q}{G} = \frac{\int_{0}^{1} \rmd z ~z~\Kc_\qg(z)}{\int_{0}^{1} \rmd z~z~K_\gq(z)} \approx 0.07\, \times 2 N_f \stackrel{N_f=3}{\approx} 0.42\;.
\end{eqnarray} 

We emphasize that the existence of these Kolmogorov type solutions does not depend on the detailed properties of the splitting Kernel. However, it does rely on the fact that the branching rates scale depend parametrically as $1/\sqrt{x}$ on the momentum fraction of the emitter. 
In order to ensure an exact cancellation of gain and loss terms in the stationarity conditions in Eqs.~(\ref{eq:statcon}), we assumed that the $x$ range of the power spectra (\ref{eq:qg-spect}) extends all the way to infinity. Even though this renders these formal solution unphysical, as it requires an infinite amount of energy, i.e., $\int_0^\infty \rmd x\, D_\glu(x) +D_\S(x) = \infty$, the KZ spectra (\ref{eq:qg-spect}) can still be relevant, as it is the case in many examples of weak-wave turbulence \cite{Nazarenko}. Specifically, we will now demonstrate that the KZ solution in Eq.~(\ref{eq:qg-spect}) is associated with a scale invariant energy flux $\dot{\epsilon}\neq 0$, from the large $x$ to the small $x$ region. Even though energy can in principle be transferred directly from highly energetic large $x$ to less energetic small $x$ partons via highly asymmetric splittings, it turns out that the interactions on the KZ spectrum are effectively local in energy, that is, they are not sensitive to the scales where energy is injected or removed. Consequently, the KZ solution can be realized approximately within an inertial range of momentum fractions $x$, e.g. in a driven/forced cascade where energy is steadily injected into the system or in freely decaying turbulence as it is the case in the present work. 


In order to establish these features, we follow the standard analysis in the context of wave turbulence \cite{Nazarenko}. We limit the support of the distribution to the physical interval $x\in (0,1]$ and consider the flux of energy below a scale $x_0$
\begin{eqnarray}
\dot{\epsilon}(x_0)=\int_{x_0}^{1}\rmd x~ \left[\partial_{\tau}D_\glu(x) + \partial_{\tau}D_\S(x)\right]\,.
\end{eqnarray}
Separating the contributions into quark and gluon initiated processes, one finds that for the gluon initiated processes $g\to gg$ and $g\to q\bar{q}$ the energy flux is given by \cite{Blaizot:2015jea,Blaizot:2013hx,Baier:2000sb}
\begin{eqnarray}
\dot{\epsilon}_\glu(x_0)&=&+\int_{x_0}^{1}\rmd x \int_{x}^{1} \rmd z~\Big( \Kc_\glug(z) + \Kc_\qg(z) \Big) \sqrt{\frac{z}{x}}D_\glu\left(\frac{x}{z}\right)  \nn
&&- \int_{x_0}^{1}\rmd x \int_{0}^{1} \rmd z~\Big( \Kc_\glug (z) + \Kc_\qg(z) \Big) \frac{z}{\sqrt{x}}D_\glu\left(x\right), 
\end{eqnarray}
where the first term corresponds to the production of a quark/anti-quark or gluon with momentum fraction $x$ resulting from the splitting of a parent gluon with momentum fraction $x/z$, whilst the second term, which is negative, corresponds to the loss of a gluon $x$ by splitting into softer fragments.

Upon changing the order of integrations and performing a change of variables to $x \to x/z$ to combine the gain and loss terms, one can express the integral as
\begin{eqnarray}
\dot{\epsilon}_\glu(x_0)&=& -\int_{x_0}^{1} \rmd z\,z\,\Big(  \Kc_\glug(z) + \Kc_\qg(z) \Big) \int_{x_0}^{x_0/z} \rmd x \,\frac{D_\glu^{s}(x)}{\sqrt{x}} \nn
	&& + \int_{0}^{x_0} \rmd z\,z\,\Big(  \Kc_\glug(z) + \Kc_\qg(z)\Big) \int_{x_0}^{1} \rmd x  \,\frac{D_\glu(x)}{\sqrt{x}}\;.
\end{eqnarray} 
Using the explicit form of the Kolmogorov spectrum, the integrals can be evaluated, and it follows that for sufficiently small values of $x_0 \ll 1$ (i.e. within the inertial range) the energy flux is scale independent  $\dot{\epsilon}_\glu(x_0\ll 1)\simeq - \gamma_\glu \, G$ with
\begin{eqnarray}\label{eq:gamma-glu}
\gamma_\glu= \int_{0}^{1} \rmd z z\Big( \Kc_\glug(z) + \Kc_\qg(z) \Big) \log\left(\frac{1}{z}\right) \approx 25.78 + 0.177 \,N_f \;. 
\end{eqnarray}
Similarly for quark initiated process $q\to gq$, the energy flux is given by
\begin{eqnarray}
\dot{\epsilon}_\q(x_0)&=&+ \int_{x_0}^{1}\rmd x \int_{x}^{1} \rmd z  \Big( \Kc_\qq(z) + \Kc_\gq(z)\Big) \sqrt{\frac{z}{x}} D_\S\left(\frac{x}{z}\right) \nn
&& -  \int_{x_0}^{1}\rmd x  \int_{0}^{1} \rmd z  \, \Kc_\qq(z)\,  \frac{1}{\sqrt{x}} D_\S(x)
\end{eqnarray} 
which, following similar steps, and using the symmetries of the kernel $\Kc_\qq(z) = \Kc_\gq(1-z)$, to re-express
\begin{eqnarray}
\int_{0}^{1} \rmd z~\Kc_\qq(z)  = \int_{0}^{1} dz \Big(z \Kc_\qq(z) + (1-z)   \Kc_\qq(z) \Big)=  \int_{0}^{1} \rmd z\,z\, \Big(\Kc_\qq(z) + \Kc_\gq(z)  \Big) \nonumber \\
\end{eqnarray}
can be evaluated in the same fashion. In the limit $x_0 \ll 1$, one again obtains a scale independent flux $\dot{\epsilon}_\q(x_0\ll 1)\simeq -\gamma_\q \,Q $ 
with the flux constant given by
\begin{eqnarray}
\gamma_\q=\int_{0}^{1} \rmd z \, z\, \Big(\Kc_\qq(z) + \Kc_\gq(z)  \Big) \log\left(\frac{1}{z}\right) \approx 11.595
\end{eqnarray}
Collecting everything the KZ spectra in \eqn{eq:qg-spect} are associated with a scale invariant energy flux
\begin{eqnarray}\label{eq:KZ-flux-qg}
\dot{\epsilon} \simeq -\gamma_\glu G -\gamma_{\q} Q \simeq   -(25.78 + 0.177 N_f) G - 11.59\, Q   \;,
\end{eqnarray}
characterizing the energy transfer from larger to smaller momentum fractions. By taking into account the quark/gluon suppression factor $Q/G\approx 0.14 N_f$ in the stationary turbulent regime (c.f. \eqn{eq:qg-ratio}), we find that for $N_f =3$ the contributions to energy flux from gluon and quark initiated processes are
\begin{eqnarray}\label{eq:KZ-flux-qg-2}
\dot{\epsilon}  \stackrel{N_f=3}{\simeq} -(\underbrace{25.8 }_{g\to gg} + \underbrace{0.5}_{g \to q\bar{q}} + \underbrace{4.9  }_{q \to gq} ) \, G
\end{eqnarray}
One therefore concludes, that the energy transfer in \eqn{eq:KZ-flux-qg} is dominated by the $g\to gg$ process, while the $q\to gq$ and $g \to q\bar{q}$ processes contribute $16\%$ and respectively $0.6\%$  to the overall result.

Since in contrast to the quark + antiquark distribution $D_\S$, the evolution of the non-singlet (valence) distribution $D_\NS^{(i)}$ does not couple to the $g\to q\bar{q}$ process, the stationary turbulent solution exhibits a different power spectrum
\begin{eqnarray}
D_\NS^{(i)}(x)=V^{(i)} \sqrt{x}\, ,
\end{eqnarray}
which can be associated with a particle cascade. Similar to the previous discussion, one can determine the particle flux of the non-singlet valence component 
\begin{eqnarray}
\dot{n}_{(i)}(x_0) = \int^{1}_{x_0} \frac{\rmd x}{x}~\partial_{\tau}D_\NS^{(i)}(x)
\end{eqnarray}
as
\begin{eqnarray}
\dot{n}_{(i)}(x_0) &=& \int_{x_0}^{1}\frac{\rmd x}{x} \int_{x}^{1} \rmd z \, \Kc_\qq(z)  \sqrt{\frac{z}{x}} D_\NS^{(i)}\left(\frac{x}{z}\right) -  
 \int_{x_0}^{1}\frac{\rmd x}{x} \int_{0}^{1} \rmd z \, \Kc_\qq(z) \, \frac{1}{\sqrt{x}} D_\NS^{(i)}(x ) \nn
 &=& -\int_{x_0}^{1} \rmd z \, \Kc_\qq(z) \int_{x_0}^{x_0/z} \rmd x  \frac{D_\NS^{(i)} (x )}{x^{3/2}} + \int_{0}^{x_0} \rmd z \, \Kc_\qq(z) \int_{x_0}^{1} \rmd x  \,\frac{D_\NS^{(i)} (x)}{x^{3/2}}
\end{eqnarray}
such that for $x_0 \ll 1$ the particle flux becomes scale independent and explicitly given by
\begin{eqnarray}
\label{eq:valence-flux}
\dot{n}_{(i)}(x_0 \ll 1) &\simeq &  -V^{(i)} \, \gamma_\NS \;, \qquad  \gamma_\NS=\int_{0}^{1} \rmd z \, \Kc_\qq(z) \log\left(\frac{1}{z}\right) \approx 6.82\;, 
\end{eqnarray}
characterizing the transport of the valence flavor from larger to smaller momentum fractions.

One important feature of our energy/particle flux analysis is that the existence of the ``boundary" at $x=1$ does not affect the scale independence of the flux for $x_0 \ll 1$. In the study of weak-wave turbulence, this feature is known as ``locality of interactions"  and plays an important role in determining the relevance of the Kolmogorov scaling solution for real world problems. Since locality of interactions guarantees that the physics near the boundary $x\simeq 1$ does not have a direct impact on the dynamics at scales $x\ll1$, we can expect that regardless of the dynamics of hard $(x\simeq 1)$ modes, the system will form a turbulent cascade within an inertial range of momenta at small $x$. Strikingly, this feature has been demonstrated explicitly in analytical solutions of purely gluonic models~\cite{Blaizot:2013hx}, where for instance for a simplified splitting kernel, the exact solution of the initial value problem takes the form
\begin{eqnarray}
D_{\rm simple}(x,\tau) = \frac{\tau }{\sqrt{x} (1-x)^{3/2}} \exp\left(-\frac{\pi\tau^2}{1-x}\right),\qquad \text{for}  \qquad \Kc_{\rm simple}(z)=\frac{1}{z^{3/2}(1-z)^{3/2}}, \nonumber \\
\end{eqnarray}
featuring the same turbulent $1/\sqrt{x}$ dependence at small-x, however with a time dependent amplitude $\tau\exp\left(-\pi\tau^2\right)$ characterizing the decay of the jet. We will discuss shortly how including quarks and gluons in the problem of jet energy loss in a thermal medium, the locality of interactions leads to the emergence of a ``decaying turbulence" with the same universal quark/gluon ratio at small $x$ predicted from the analysis of the stationary case. Besides the real-world example of jet energy loss, we also provide a brief discussion of the properties of stationary turbulent solutions for driven/forced cascades in App.~\ref{sec:AppendixStationaryTurbulence}.

\section{Quenching of Quark \& Gluon jets}
\label{sec:Results}
We will now discuss the quenching dynamics of quark and gluon jets based on numerical solutions of the kinetic equations (\ref{eq:evol-eq-glu},\ref{eq:evol-eq-S},\ref{eq:evol-eq-NS}). We will focus on the evolution of the in-medium fragmentation functions $D(x,\tau)$, and distinguish between gluon and quark fragments inside the jet. 

\begin{figure}[ht!]
\begin{center}
\begin{minipage}{0.475\textwidth}
\includegraphics[width=\textwidth]{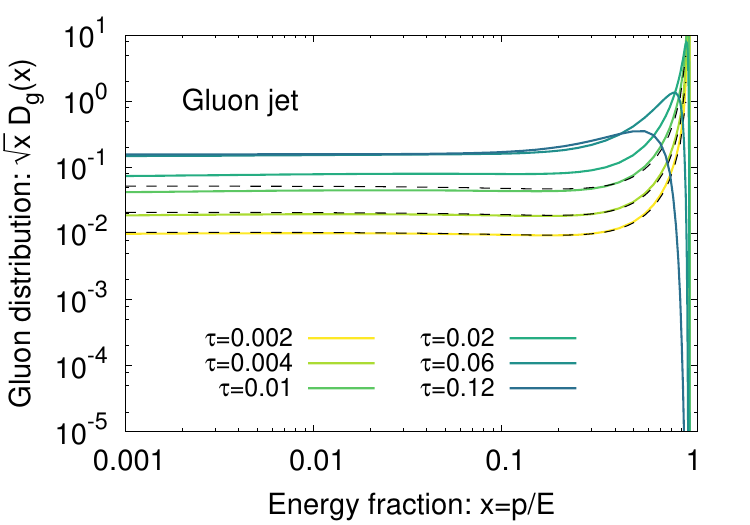}
\end{minipage}
\begin{minipage}{0.475\textwidth}
\includegraphics[width=\textwidth]{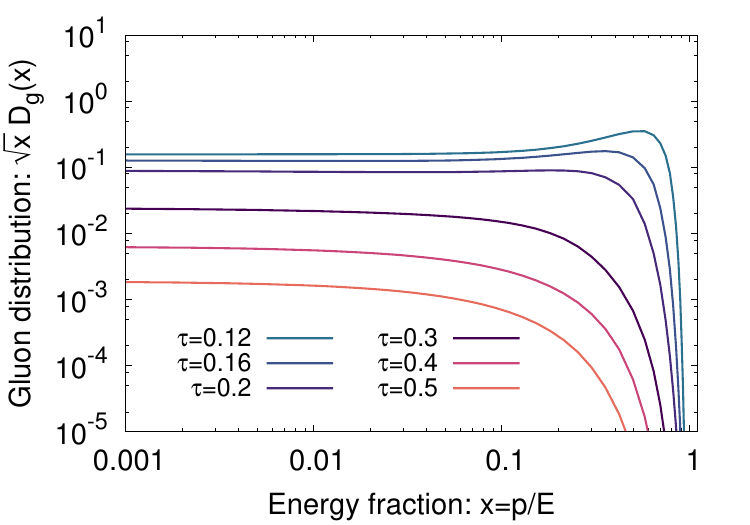}
\end{minipage}

\begin{minipage}{0.475\textwidth}
\includegraphics[width=\textwidth]{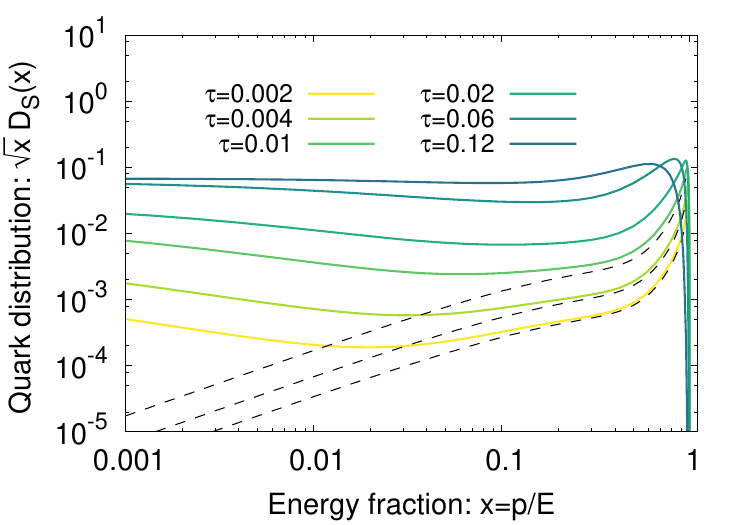}
\end{minipage}
\begin{minipage}{0.475\textwidth}
\includegraphics[width=\textwidth]{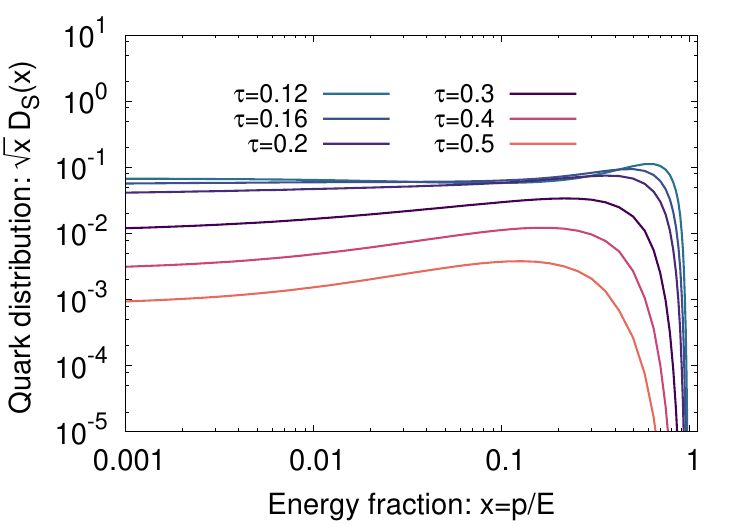}
\end{minipage}
\caption{\label{fig:GluonJet} Evolution of the quark/gluon fragmentation functions $\sqrt{x} D_{\glu}(x)$ (top) and $\sqrt{x}D_{\S}(x)$ (bottom) for \emph{gluon jets} as a function of energy fraction $x$. Different curves in the top and bottom panels show the results for different evolution times $\tau$ corresponding to different amounts of energy loss in the medium. Dashed curves in the left panels show the result for a single $g\to gg$ or $g\to q\bar{q}$ branching. See text for further discussion.}
\end{center}
\end{figure}

Our results for the radiative break up of a gluon jet are compactly summarized in Fig.~\ref{fig:GluonJet}. Different panels show the distributions $\sqrt{x}D_{\glu/\S}(x)$ of gluon (top) and quark (bottom) fragments, at various early (left panels) and late times (right panels) of the evolution. Starting at early times $\tau \sim 10^{-3}$, one observes how the soft fragments with $x \ll 1$ are radiated from the original hard parton. Clearly, at intermediate values of $x$ the radiative spectrum initially follows the perturbative shape (c.f. Eqn~(\ref{eq:lo-glu})) indicated by the black dashed lines and characterized by an approximate scaling $D_\glu(x)\propto x^{-1/2}$ for gluons and $D_\S(x)\propto  x^{-3/2}$ for quarks. However, at small $x$ clear deviations from the perturbative spectrum emerge already at such early times, as is visible most prominently for the quark distribution. Since the probability $P_{\rm split} \sim \tau/\sqrt{x}$ to undergo subsequent splittings, becomes of order one at a critical value $x_{\br} \sim \tau^2$, the distributions in this small $x$ regime are dominated by multiple branchings, such that for example processes where a soft quark emits another gluon or a soft quark/anti-quark pair is produced from a soft gluon play an important role in determining the spectrum. 

\begin{figure}[ht!]
\begin{center}
\includegraphics[width=0.7\textwidth]{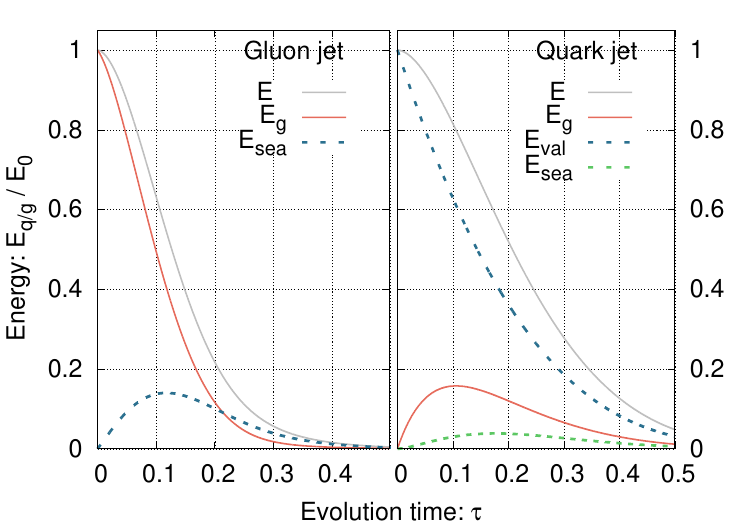}
\includegraphics[width=0.7\textwidth]{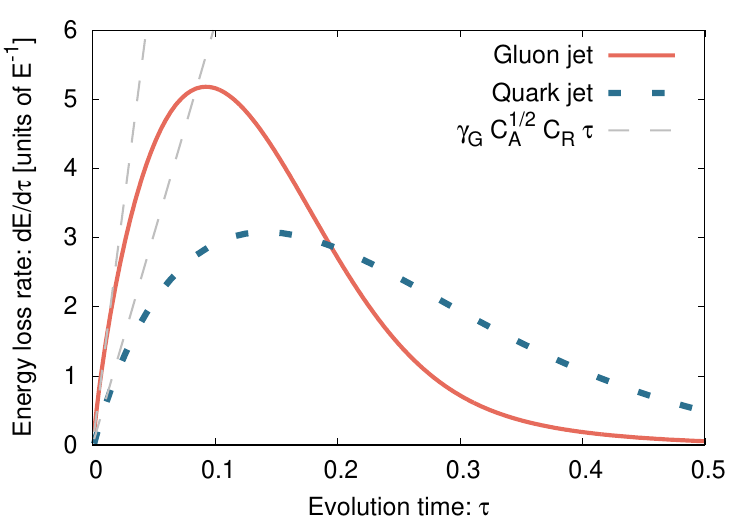}
\caption{\label{fig:EnergyJet} (left) Evolution of the energy $E_{\rm g/sea/val}$ carried by gluons (red solid), sea (blue dashed) and valence quarks (green dashed)  and inside a quark/gluon jet. (right) Differential energy loss $\rmd E/\rmd \tau$ for quark and gluon jets.}
\end{center}
\end{figure}

By following the evolution to later times, one observes that the characteristic scale $x_\br$ where multiple emissions become important increases towards larger $x$ values. Quark and gluon distributions at small $x$ start to show an approximate $1/\sqrt{x}$ power law dependence, which is fully developed up to $x\sim0.1$ by the time $\tau\sim0.1$. Since the emergence of the turbulent $\sim 1/\sqrt{x}$ spectrum can be associated with an energy flux from the large $x\sim1$ to the small $x$ region, significant changes in the large $x$ part of the spectrum also start to take place on the same time scale, depleting the original peak around $x\sim1$. Since the energy flux becomes scale invariant at small $x$, energy lost by hard fragments is transported all the way to $x\sim T/E$ where it is absorbed by the thermal medium. By the time $\tau\sim0.1$ the jet as whole has lost $\sim40\%$ of its initial energy to the thermal medium (c.f. Fig.~\ref{fig:EnergyJet}). 

Beyond $\tau\sim 0.1$ the peak at large $x$ begins to disappear as the radiative break-up processes enters a regime of decaying turbulence. Successively, all large $x$ fragments disappear, while the small $x$ part of the quark and gluon distributions continue to follow the turbulent $\sim1/\sqrt{x}$ behavior throughout the evolution. Interestingly, one finds that at late times $\tau \gtrsim 0.2$ quarks begin to dominate the large $x$ part of the distribution. Even though the jet has already lost $\approx 80\%$ of its energy by this time, and the probability 
\begin{eqnarray}
P_{>}(x)=\frac{\int_{x}^{1} dz~D_\glu(z)+D_\S(z) }{\int_{0}^{1}dz~D_\glu(z)+D_\S(z)}
\end{eqnarray}
to find fragments with $x>0.3$ is less than $30\%$ in this regime, it is is in fact more likely that such a large $x$ fragment is a quark/anti-quark rather than a gluon.\\


\begin{figure}[ht!]
\begin{center}
\begin{minipage}{0.475\textwidth}
\includegraphics[width=\textwidth]{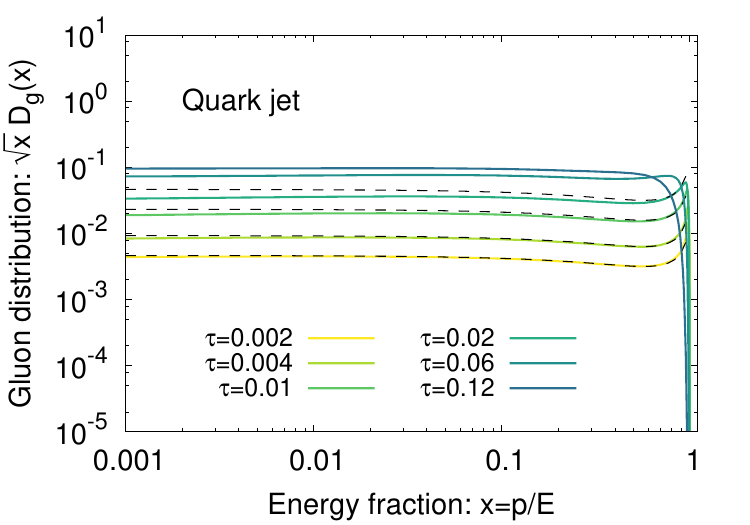}
\end{minipage}
\begin{minipage}{0.475\textwidth}
\includegraphics[width=\textwidth]{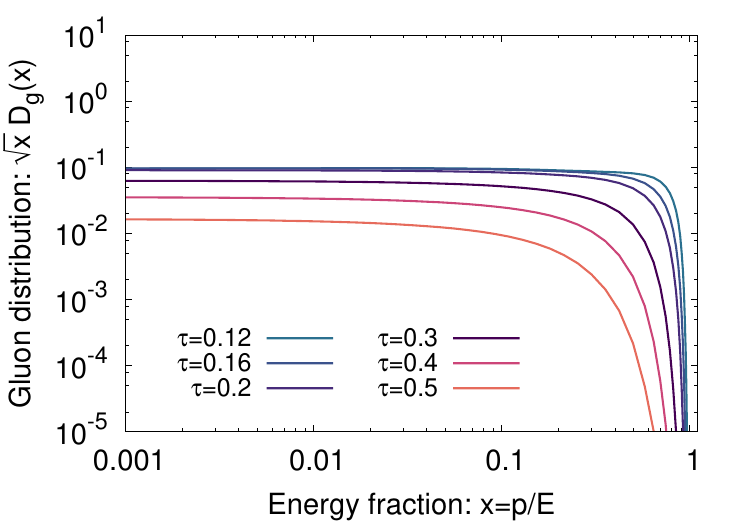}
\end{minipage}

\begin{minipage}{0.475\textwidth}
\includegraphics[width=\textwidth]{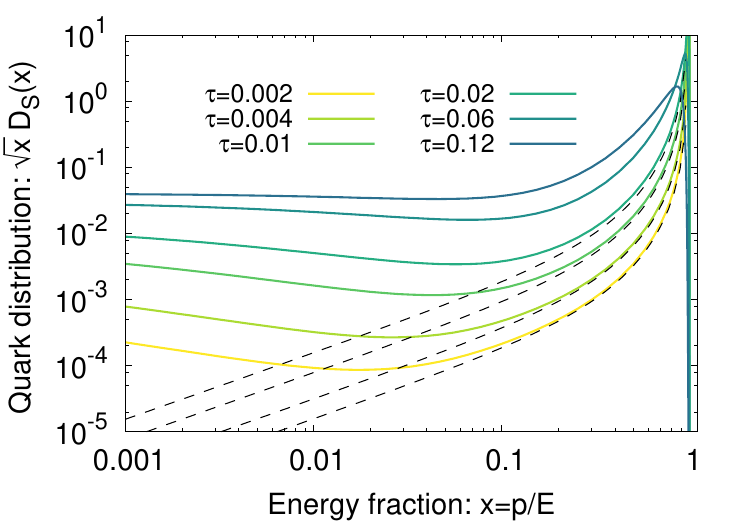}
\end{minipage}
\begin{minipage}{0.475\textwidth}
\includegraphics[width=\textwidth]{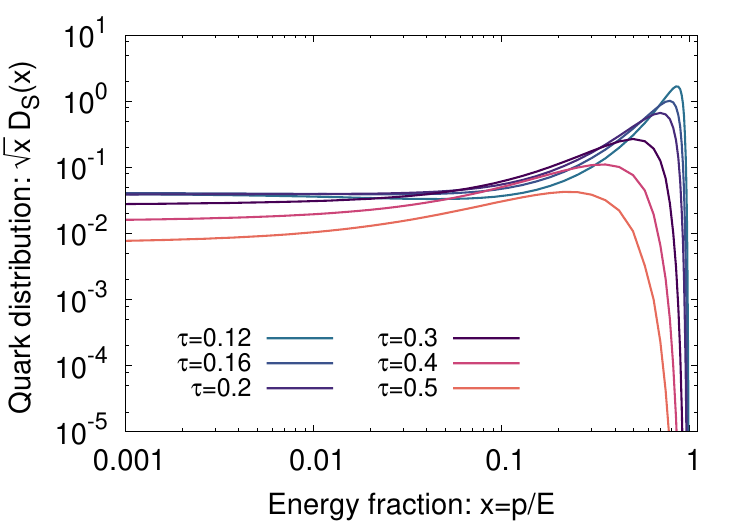}
\end{minipage}
\begin{minipage}{0.475\textwidth}
\includegraphics[width=\textwidth]{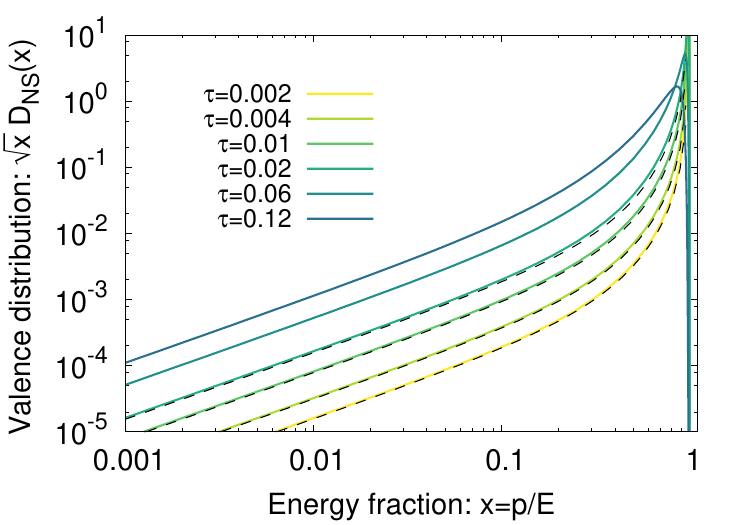}
\end{minipage}
\begin{minipage}{0.475\textwidth}
\includegraphics[width=\textwidth]{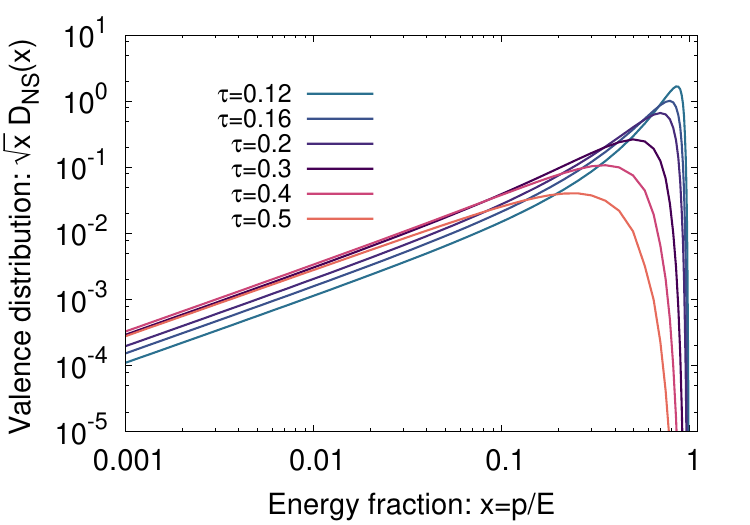}
\end{minipage}
\caption{\label{fig:QuarkJet} Evolution of the quark/gluon fragmentation functions $\sqrt{x} D_\glu(x)$ (top), $\sqrt{x}D_\S(x)$ (middle) and $\sqrt{x} D_\NS(x)$ (bottom)  for \emph{quark jets} as a function of energy fraction $x$. Different curves in each panels show the results for different evolution times $\tau$ corresponding to different amounts of energy loss in the medium. Dashed curves in the left panels show the result for a single $q\to qg$ branching. See text for further discussion.}
\end{center}
\end{figure}

Similar observations can be made for the evolution of the gluon and (flavor singlet) quark distributions inside a quark jet, which are presented in the top and middle panels of Fig.~\ref{fig:QuarkJet}. In accordance with our previous discussion, one finds that multiple branchings quickly establish a turbulent $1/\sqrt{x}$ power spectrum at small $x$, which persists over the coarse of the entire evolution. Once the hard fragments undergo a quasi-democratic splitting with order one probability (around $\tau \sim 0.15$), the radiative break-up process enters the regime of decaying turbulence characterized by an (inverse) energy cascade moving towards smaller $x$. Interestingly, one observes that at very late times $\tau \sim 0.5$ the fragmentation spectra of quark and gluon jets shown in Figs.~\ref{fig:GluonJet} and \ref{fig:QuarkJet} become rather similar, indicating an effective memory loss of the initial conditions. 

One additional feature of quark jets concerns the evolution of the (flavor non-singlet) valence distribution $\sqrt{x} D_\NS(x)$ shown in the bottom panels of Fig.~\ref{fig:QuarkJet}. Starting from early times $\tau \lesssim 10^{-2}$, the emission of gluon radiation off the valence charge leads to an increase of the valence distribution for $x<1$, and the distribution is well described by the perturbative solution as indicated by black dashed lines. Even though the perturbative solution fails to describe the evolution beyond early times, the characteristic $x^{-3/2}$ spectrum at small $x$ clearly remains present throughout the entire evolution. However, as shown explicitly in \cite{Blaizot:2015jea}, this apparent agreement with the perturbative power spectrum is purely accidental. In accordance with our discussion in Sec.~\ref{sec:Analytics}, the emergence of a  $x^{-3/2}$ power spectrum at late times should be associated with the scale invariant flux of the valence particle number to smaller $x$, such that by $\tau=0.3,(0.5)$ there is about a $30 (70)\%$ probability that the valence charge has been lost to the thermal medium.

\subsection{Energy loss of quark and gluon jets}
Based on our results for the in-medium fragmentation functions, we will now analyze the energy loss of quark and gluon jets in more detail. Our results are compactly summarized in the left panel of Fig.~\ref{fig:EnergyJet}, where different curves labeled $E$ and $E_{\rm g/val/sea}$ show the total energy of all jet fragments as well as the individual energy fractions 
\begin{eqnarray}
E_\glu=\int \rmd x~D_\glu(x)\;, \qquad E_{\rm sea}=\int \rmd x~\Big(D_\S(x)-D_\NS(x)\Big)\;, \qquad E_{\rm val}=\int \rmd x~D_\NS(x)\;,
\end{eqnarray}
carried by gluons, sea and valence quark fragments of the jet, normalized to the initial jet energy $E_0$ as a function of time. We first note that the turbulent cascade creates an energy flux to arbitrarily small $x$, such that a finite amount of energy $\Delta E$ is lost, irrespective of the cutoff scale distinguishing between the thermal medium and the soft fragments of the jet. Concerning the chemical composition, one finds that the energy loss of the original parton (gluon/quark) is accompanied by an increase of the energy carried by quarks (gluons) inside a gluon (quark) jet, which reaches a maximum around $\tau \sim 0.1$ when approximately $15\%$ of the energy is carried by the opposite species.  However, on large time scales, regardless of the chemical composition of the initial jet, the dominant energy fraction is always carried by the quark degrees of freedom, as the large $x$ quarks tend to loose their energy more slowly as compared to the large $x$ gluons. Specifically for quark jets, the large $x$ valence contribution $E_{\rm val}$ dominates the energy throughout the entire evolution. However, even for gluon jets, the (flavor singlet) quark component starts to dominate the energy around times $\tau=0.2$. Despite the fact that the jet has already lost about $\sim80\%$ of its initial energy by this time, the remaining energy should still be sufficient to detect the jet and it would be interesting to investigate possible experimental signatures of this ``medium filtering" mechanism. 

Besides the total energy, it is also interesting to analyze the differential energy loss rates $\rmd E/\rmd \tau$ shown in the right panel of Fig.~\ref{fig:EnergyJet}, where we compare our results for quark and gluon jets. Notably at very early times $\tau\lesssim 0.02$, the energy loss can be understood quantitatively based on the following considerations. By emission of soft gluon radiation from the original hard parton, the jet establishes a $G/\sqrt{x}$ gluon spectrum at small $x$ with a linearly increasing amplitude $G\approx \tau C_{A}^{1/2} C_{R}$ at early times (cf. Eq.~(\ref{eq:lo-glu-smallx})). Since the additional radiation emitted by these soft  $(x \ll 1)$ gluons induces a finite scale invariant energy flux towards arbitrarily small values of $x$, the jet as a whole looses energy to the thermal bath. Combining the (perturbative) estimate of $G$ with the associated energy flux $\dot{\epsilon}\approx\gamma_{\glu} G$ (c.f. Eq.~(\ref{eq:gamma-glu})) the energy loss rate at early times can be determined as 
\begin{eqnarray}\label{eq:dEdxEarly}
\left. \frac{1}{E} \frac{\rmd E}{\rmd \tau} \right|_{\tau \ll 1} \approx \gamma_\glu  \, C_A^{1/2} C_R\,  \tau,
\end{eqnarray}
such that in terms of the original variables $\frac{dE}{dt} = \bar{\alpha}^2 \hat{\bar{q}} C_{A}^{1/2} C_{R} \gamma_{\glu} t$ which is illustrated by the (gray) dashed lines in Fig.~\ref{fig:EnergyJet}.  One observes that the estimate in Eq.~(\ref{eq:dEdxEarly}) provides an accurate description of the energy loss for  $\tau\lesssim 0.02$. Based on Eq.~(\ref{eq:dEdxEarly}) one concludes that in this regime, the energy loss of quark and gluon jets are related by simple Casimir scaling
\begin{eqnarray}
\left. \frac{\rmd E}{\rmd \tau}\right|^{\text{Quark jet}}_{\tau \ll 1}  = C_{F}/C_{A}\left. \frac{\rmd E}{\rmd \tau}\right|^{\text{Gluon jet}}_{\tau \ll 1} \;.
\end{eqnarray}
However, a simple relation of this form fails to describe the energy loss at late times, where the chemistry of jet fragments is strongly altered. Instead one observes from Fig.~\ref{fig:EnergyJet} that the time scales for the energy loss of quark and gluon jets are inherently different and determined by the dynamics of multiple branching processes.

\begin{figure}[ht!]
\begin{center}
\includegraphics[width=\textwidth]{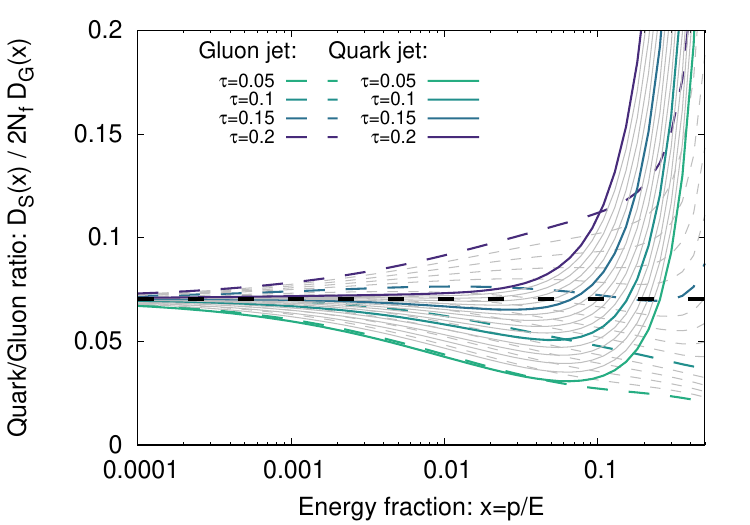}
\caption{\label{fig:QGRatio} Quark/Gluon ratio $D_\S(x)/2N_fD_\glu(x)$ of jet fragments as a function of the momentum fractions $x=p/E$. Different curves show the results for different evolution times $\tau$ of quark (solid lines) and gluon jets (dashed lines). Over an inertial range of momentum fractions $T/E \ll x \ll 1$, the ratio $D_\S/D_\glu$ of quark to gluon fragments is identical for quark and gluon jets and close to the Kolmogorov ratio (c.f. Eq.~(\ref{eq:qg-ratio})) shown by a black dashed line.}
\end{center}
\end{figure}

\subsection{Chemistry of jet fragments}
One of the most striking features of the turbulent jet energy loss mechanism is the universal quark to gluon ratio for soft fragments of the jet derived in Sec.~\ref{sec:Analytics}. In order to investigate, to what extent this feature is born out in the radiative break-up of quark and gluon jets, we present in Fig.~\ref{fig:QGRatio} the ratio of in-medium fragmentation functions  $D_\S(x)/2N_fD_\glu(x)$. Solid (dashed) lines  in Fig.~\ref{fig:QGRatio} show the results for (quark) gluon jets plotted as a function of the momentum fraction $x$ at various different times of the evolution. One observes that over a large range of evolution times -- corresponding to different amounts of jet energy loss -- and momentum fractions $x$ the ratio of quarks to gluon fragments is close to the universal Kolmogorov ratio $D_\S/D_\glu \approx 0.07 \times 2N_f$, derived in Eq.~(\ref{eq:qg-ratio}) and shown by a black dashed line. Even at rather early and rather late times, one finds that the ratio $D_\S/D_\glu$ approaches the universal value at very small $x\lesssim 10^{-3}$, where the chemical composition is determined by the local balance of the $g\to q\bar{q}$ and $q\to gq$ processes. However, considering the importance of additional processes at scales $x\sim T/E$, it seems unrealistic that under typical experimental conditions where 
$E_{\rm jet} \sim 100~{\rm GeV}$ and $T\sim 300~{\rm MeV}$, a separation of scales $T/E \gtrsim 10^{3}$ can be achieved. Nevertheless, our results in Fig.~\ref{fig:QGRatio} clearly suggest the existence of a wide enough kinematic range ($0.2\gtrsim \Delta E/E \lesssim0.8$, $x\lesssim 0.1$) where experimental signatures of the modified flavor composition may be explored.

\section{Conclusions \& Outlook}
\label{sec:Conclusions}
Based on a coupled set of evolution equations, describing the $g\to gg$,$g\to q\bar{q}$ and $q\to qg$ branching processes of (on-shell) quarks and gluons, we have analyzed the in-medium jet fragmentation of quark and gluon jets, with an emphasis on the chemical composition of jet fragments. Our findings can be compactly summarized as follows:
\begin{itemize}
\item Energy loss of a highly energetic jet to a thermal medium is realized via a turbulent cascade, associated with a scale independent energy flux from momentum scales on the order of the jet energy $p\sim E$ all the way to the energy scale of the medium $p \sim T$
\item Since splitting rates of soft $(x\ll 1)$ fragments are enhanced by a factor $1/\sqrt{x}$ relative to the splitting rates of the original hard parton, multiple branchings determine the properties of the in-medium fragmentation function at small $x$, where the spectrum reaches a non-equilibrium steady state characterized by a $1/\sqrt{x}$ power law dependence of the quark and gluon fragmentation functions. 
\item Since splittings are sufficiently local in momentum space (c.f. Sec.~\ref{sec:Analytics}), the spectrum at small $x$ is insensitive to the large $x$ structure of the jet except for the energy transmitted by the large $x$ fragments.  Most strikingly, the chemical composition of small $x$ jet fragments is entirely determined by the balance of the $g\to q\bar{q}$ and $q\to gq$ splitting processes and given by $\frac{D_{\S}(x)}{D_{\glu}(x)}= \frac{\int_{0}^{1} \rmd z~z~K_\qg(z)}{\int_{0}^{1} \rmd z ~z~\Kc_\gq(z)} \approx 0.07 \times 2N_f$ to leading logarithmic accuracy. 
\item Energy loss of quark and gluon jets follows Casimir scaling (c.f. Eq.~(\ref{eq:dEdxEarly})) for short in-medium evolution times, i.e. for jets which lose a sufficiently small amount of energy. However, for jets which lose an appreciable amount of their energy Casimir scaling breaks down, as the chemistry of fragments is strongly alterexd. Energy loss rates are dominated by gluon radiation -- contributions of dynamical quarks to jet energy loss are on the $10-20\%$ level.
\item Energy loss in the medium provides an efficient filtering mechanism: for jets which exhibit a sufficiently large energy loss the large $x$ part of the in-medium fragmentation functions is always dominated by quarks. While for quark jets, the valence flavor dominates at large $x$, one finds for example that for gluon jets loosing $80\%$ of their initial energy, approximately $1/4$ of the remaining energy is carried by strange quarks.
\end{itemize}

Some of our findings regarding the chemistry of jet fragments, may have interesting signatures in high-energy heavy-ion experiments. Considering for example the universal quark/gluon ratio at small $x$ along with the fact that strange quarks are more likely to produce strange hadrons, one should expect to observe modifications of the $K/\pi$ or $\Lambda/\pi$ ratio inside jets in heavy-ion collisions, compared to the baseline of jets $p+p$ as well as the thermal strangeness contribution of the medium. Similarly, one naturally expects the medium filtering to manifest itself in terms of a strangeness enhancement of large $x$ fragments in strongly quenched jets. Of course, to provide explicit predictions of these and other effects, it will be necessary to extend our study in various regards to properly include e.g. vacuum emissions and hadronization effects. In this context, it could also be interesting to extend our study to include charm degrees of freedom to study e.g. $D$ meson production inside jets, at the expense of an additional scale related to the charm quark mass.  This is work in progress and will be reported elsewhere.

Besides offering new insights into the chemistry of jet fragmentation, our study also provides a first step towards a better understanding of the chemical evolution of the quark gluon plasma during the pre-equilibrium stage. While the initial state immediately after the collision of heavy nuclei is expected to be highly gluon dominated, more than $2/3$ of the energy of the equilibrated QGP is carried by quark degrees of freedom. Based on a seminal paper \cite{Baier:2000sb} the processes that eventually lead to the formation of an equilibrated plasma are believed to be strongly reminiscent of the process of jet-quenching, with the equilibration time determined by the time it takes for a typical (semi-hard) parton to loose all of its initial energy. Various recent works~\cite{Baier:2000sb,Berges:2013eia,Kurkela:2014tea,Kurkela:2018vqr} have consolidated the estimates of \cite{Baier:2000sb} based on sophisticated numerical simulations, however with the exception of \cite{Tanji:2017suk,Tanji:2017xiw} these studies have focused exclusively on the gluon degrees of freedom. Based on our results, one can readily conclude that a small fraction of $\approx 0.07 N_f$ quarks per gluon are produced directly as a result of the radiative break-up process. However, since this ratio is small compared to the equilibrium value of $\approx 0.75 N_f$ quarks per gluon, it is also clear that processes, such as e.g. elastic $gg\to q\bar{q}$ conversions, operative at lower momentum scales will play an important role in the chemical equilibration of the quark gluon plasma.


\section*{Acknowledgement:} We thank J.P.~Blaizot, E.~Iancu,A.~Kurkela, A.~Mazeliauskas, G.D.~Moore, J.F.~Paquet and D.~Teaney for insightful discussions and collaboration on related projects. We would also like to express our gratitude to the organizers of the 2018 Santa-Fe Jet and Heavy-Flavor Workshop, where this work was initiated. This work was supported in part by the U.S. Department of Energy, Office of Science, Office of Nuclear Physics under Award Numbers DE-FG02-97ER41014 (SS) as well as DE-FG02-00ER41132 and DE-SC0012704 (YMT).

\appendix
\section{Stationary solutions for forced cascades}
\label{sec:AppendixStationaryTurbulence}
\begin{figure}[ht!]
\begin{center}
\includegraphics[width=0.8\textwidth]{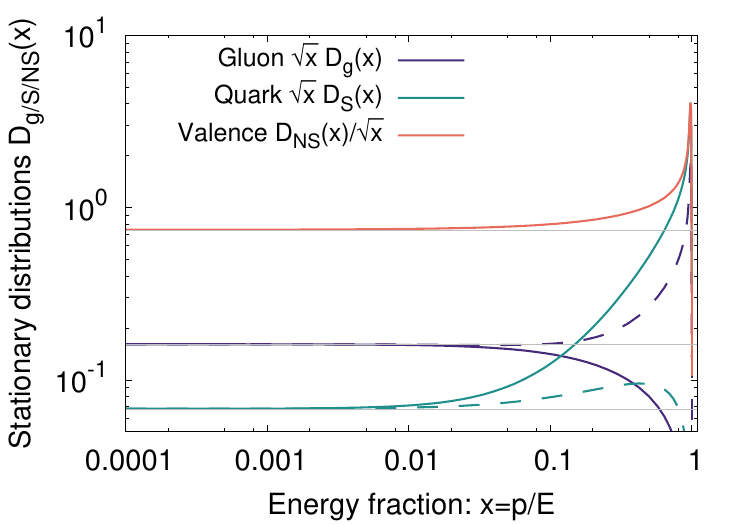}
\caption{\label{fig:StationarySol} Stationary solutions for the distributions $D_{\rm g/S/NS}(x)$ in a driven cascade. Solid curves correspond to quark forcing ($F_\glu=0\;,F_\S=5\;,F_\NS=5$) whereas dashed curves show the results for gluon forcing ($F_\glu=5\;,F_\S=0\;,F_\NS=0$). Horizontal gray lines show the KZ solution, which is realized to high accruacy in an inertial range of energy fractions $x\lesssim 0.02$.}
\end{center}
\end{figure}

Below we provide additional results for the stationary Kolmogorov solutions obtained in driven cascades. In order to realize this scenario, we simply add the following source terms to the right hand side of the evolution equations (\ref{eq:evol-eq-glu},\ref{eq:evol-eq-S},\ref{eq:evol-eq-NS}) 
\beq
F_{\glu}(x)=F_{\glu}~\delta(1-x)\;, \qquad   F_{S}(x)=F_\S~\delta(1-x)\;, \qquad  F_{\glu}(x)=F_\NS~\delta(1-x)\;,  
\eeq 
which inject energy and valence particle number at $x=1$ at constant rates
\begin{eqnarray}
\label{eq:EDotInj}
\dot{\epsilon}=F_{\glu}+F_{\S}\;, \qquad \dot{n}=F_{\NS}\;.
\end{eqnarray}
We then solve the evolution equations for the fragmentation functions in the presence of the source terms until the system relaxes to the stationary solution. Numerical results for the stationary solutions are presented in Fig.~\ref{fig:StationarySol}. \\

Based on our discussion in Sec.~\ref{sec:KolmogorovSolution}, we expect the stationary spectrum to take the form
\begin{eqnarray}
D_\glu(x)=  \frac{G}{\sqrt{x}}\;, \qquad  D_\S(x)=  \frac{Q}{\sqrt{x}}\;, \qquad D_\NS(x)=  V \sqrt{x}\;, 
\end{eqnarray}
within in an inertial range of momentum fractions $x\ll 1$. Since the energy and particle injection rates are fixed according to Eq.~(\ref{eq:EDotInj}), the stationary amplitudes $G,Q,V$ can be determined directly by matching the energy and particle flux in Eqns.~(\ref{eq:KZ-flux-qg},\ref{eq:valence-flux}) and (\ref{eq:EDotInj}), yielding
\begin{eqnarray}
G=\frac{\dot{\epsilon}}{\gamma_\glu + \gamma_{\rm q} Q/G}\;, \qquad  V=\frac{\dot{n}}{\gamma_{NS}}\;,
\end{eqnarray}
with the universal $Q/G$ ratio as in Eq.~(\ref{eq:qg-ratio}). Specifically for $N_f=3$ light flavors, one obtains the relations
\begin{eqnarray}
G\simeq 0.032 (F_{\glu}+F_{\S})\;, \qquad Q= 0.013 (F_{\glu}+F_{\S})\;, \qquad  V=0.146 F_{\NS}\;,
\end{eqnarray}
indicated by horizontal gray lines in Fig.~\ref{fig:StationarySol}. By comparing the the full numerical solution with the KZ solution in the scaling regime, one concludes that the inertial range of the energy and particle cascades extends approximately to energy fractions $x\lesssim 0.02$. Above this scale the influence of the source on the distribution becomes clearly visible.

%



\bibliographystyle{elsarticle-num}
\bibliography{jetquenching}

\begin{thebibliography}{9}
%

\bibitem{Bjorken:1982}
J.~D.~Bjorken, 
Fermilab-Pub-82/59-THY, Batavia (1982);
Erratum, unpublished

\bibitem{Adcox:2001jp} 
  K.~Adcox {\it et al.} [PHENIX Collaboration],
  Phys.\ Rev.\ Lett.\  {\bf 88}, 022301 (2002)
  doi:10.1103/PhysRevLett.88.022301
  [nucl-ex/0109003].
  \bibitem{Adler:2002xw} 
  C.~Adler {\it et al.} [STAR Collaboration],
  Phys.\ Rev.\ Lett.\  {\bf 89}, 202301 (2002)
  doi:10.1103/PhysRevLett.89.202301
  [nucl-ex/0206011].
\bibitem{Aad:2010bu} 
  G.~Aad {\it et al.} [ATLAS Collaboration],
  Phys.\ Rev.\ Lett.\  {\bf 105}, 252303 (2010)
  doi:10.1103/PhysRevLett.105.252303
  [arXiv:1011.6182 [hep-ex]].
\bibitem{Chatrchyan:2012nia} 
  S.~Chatrchyan {\it et al.} [CMS Collaboration],
  Phys.\ Lett.\ B {\bf 712}, 176 (2012)
  doi:10.1016/j.physletb.2012.04.058
  [arXiv:1202.5022 [nucl-ex]].
\bibitem{Aad:2014bxa} 
  G.~Aad {\it et al.} [ATLAS Collaboration],
  Phys.\ Rev.\ Lett.\  {\bf 114}, no. 7, 072302 (2015)
  doi:10.1103/PhysRevLett.114.072302
  [arXiv:1411.2357 [hep-ex]].
  
\bibitem{Abelev:2013kqa} 
  B.~Abelev {\it et al.} [ALICE Collaboration],
  JHEP {\bf 1403}, 013 (2014)
  doi:10.1007/JHEP03(2014)013
  [arXiv:1311.0633 [nucl-ex]].
\bibitem{Ali:2010tw} 
  A.~Ali and G.~Kramer,
  Eur.\ Phys.\ J.\ H {\bf 36}, 245 (2011)
  doi:10.1140/epjh/e2011-10047-1
  [arXiv:1012.2288 [hep-ph]].
\bibitem{Sapeta:2015gee} 
  S.~Sapeta,
  Prog.\ Part.\ Nucl.\ Phys.\  {\bf 89}, 1 (2016)
  doi:10.1016/j.ppnp.2016.02.002
  [arXiv:1511.09336 [hep-ph]].
  
\bibitem{Mehtar-Tani:2013pia} 
  Y.~Mehtar-Tani, J.~G.~Milhano and K.~Tywoniuk,
  Int.\ J.\ Mod.\ Phys.\ A {\bf 28}, 1340013 (2013)
  doi:10.1142/S0217751X13400137
  [arXiv:1302.2579 [hep-ph]].
\bibitem{Blaizot:2015lma} 
  J.~P.~Blaizot and Y.~Mehtar-Tani,
  Int.\ J.\ Mod.\ Phys.\ E {\bf 24}, no. 11, 1530012 (2015)
  doi:10.1142/S021830131530012X
  [arXiv:1503.05958 [hep-ph]].
  
  Although jet evolution in the vacuum is well established theoretically \cite{Ali:2010tw,Sapeta:2015gee}, the theory of jet energy loss in the presence of a dense medium remains an active field of study (see \cite{Mehtar-Tani:2013pia,Blaizot:2015lma} for recent reviews). 

  \bibitem{Baier:2000sb}
R.~Baier, A.~H. Mueller, D.~Schiff, and D.~Son, {\it {`Bottom up'
  thermalization in heavy ion collisions}},  {\em Phys.Lett.} {\bf B502} (2001)
  51--58, [{{\tt
  hep-ph/0009237}}].
  

\bibitem{Kurkela:2018vqr} 
  A.~Kurkela, A.~Mazeliauskas, J.~F.~Paquet, S.~Schlichting and D.~Teaney,
  arXiv:1805.00961 [hep-ph].


\bibitem{Blaizot:2013hx} 
  J.~P.~Blaizot, E.~Iancu and Y.~Mehtar-Tani,
  Phys.\ Rev.\ Lett.\  {\bf 111}, 052001 (2013)
  [arXiv:1301.6102 [hep-ph]].
  

\bibitem{Blaizot:2015jea} 
  J.~P.~Blaizot and Y.~Mehtar-Tani,
  Annals Phys.\  {\bf 368}, 148 (2016)
  doi:10.1016/j.aop.2016.01.002
  [arXiv:1501.03443 [hep-ph]].

  
\bibitem{Altarelli:1977zs} 
  G.~Altarelli and G.~Parisi,
  Nucl.\ Phys.\ B {\bf 126}, 298 (1977).
  doi:10.1016/0550-3213(77)90384-4
  
  



 \bibitem{Baier:1996kr}
R.~Baier, Y.~L. Dokshitzer, A.~H. Mueller, S.~Peigne, and D.~Schiff, {\it
  {Radiative energy loss of high-energy quarks and gluons in a finite volume
  quark - gluon plasma}},  {\em Nucl.Phys.} {\bf B483} (1997) 291--320.

\bibitem{Baier:1996sk}
R.~Baier, Y.~L. Dokshitzer, A.~H. Mueller, S.~Peigne, and D.~Schiff, {\it
  {Radiative energy loss and p(T) broadening of high-energy partons in
  nuclei}},  {\em Nucl.Phys.} {\bf B484} (1997) 265--282.

\bibitem{Baier:1998kq}
R.~Baier, Y.~L. Dokshitzer, A.~H. Mueller, and D.~Schiff, {\it {Medium induced
  radiative energy loss: Equivalence between the BDMPS and Zakharov
  formalisms}},  {\em Nucl.Phys.} {\bf B531} (1998) 403--425,
  [{{\tt hep-ph/9804212}}].


\bibitem{Zakharov:1996fv}
B.~Zakharov, {\it {Fully quantum treatment of the Landau-Pomeranchuk-Migdal
  effect in QED and QCD}},  {\em JETP Lett.} {\bf 63} (1996) 952--957.


\bibitem{Wiedemann:2000za}
U.~A. Wiedemann, {\it {Gluon radiation off hard quarks in a nuclear
  environment: Opacity expansion}},  {\em Nucl.Phys.} {\bf B588} (2000)
  303--344, [{{\tt
  hep-ph/0005129}}].



\bibitem{Arnold:2002ja}
P.~B. Arnold, G.~D. Moore, and L.~G. Yaffe, {\it {Photon and gluon emission in
  relativistic plasmas}},  {\em JHEP} {\bf 0206} (2002) 030,
  [{{\tt hep-ph/0204343}}].
  
  \bibitem{Blaizot:2014ula} 
  J.~P.~Blaizot, Y.~Mehtar-Tani and M.~A.~C.~Torres,
  {\it Angular structure of the in-medium QCD cascade},
  arXiv:1407.0326 [hep-ph].

\bibitem{Kurkela:2014tla} 
  A.~Kurkela and U.~A.~Wiedemann,
  {\it Picturing perturbative parton cascades in QCD matter},
  Phys.\ Lett.\ B {\bf 740}, 172 (2014)
  [arXiv:1407.0293 [hep-ph]].
  
  \bibitem{Blaizot:2014rla} 
  J.~P.~Blaizot, L.~Fister and Y.~Mehtar-Tani,
{\it Angular distribution of medium-induced QCD cascades},
  arXiv:1409.6202 [hep-ph].
  

\bibitem{Arnold:2008zu} 
  P.~B.~Arnold and C.~Dogan,
  Phys.\ Rev.\ D {\bf 78}, 065008 (2008)
  doi:10.1103/PhysRevD.78.065008
  [arXiv:0804.3359 [hep-ph]].




  \bibitem{KKMN}
A. V. Kats, V. M. Kontorovich, S. S. Moiseev, and V. E. Novikov
, Zh. Eksp. Teor. Fiz. 71, 171 (1976)
  

  


\bibitem{Dokshitzer:1991wu} 
  Y.~L.~Dokshitzer, V.~A.~Khoze, A.~H.~Mueller and S.~I.~Troian,
  {\it Basics of perturbative QCD},
  Gif-sur-Yvette, France: Ed. Frontieres (1991) 274 p. 
  
  


\bibitem{Jeon:2003gi}
S.~Jeon and G.~D. Moore, {\it {Energy loss of leading partons in a thermal QCD
  medium}},  {\em Phys.Rev.} {\bf C71} (2005) 034901,
  [{{\tt hep-ph/0309332}}].



\bibitem{Gribov:1972ri} 
  V.~N.~Gribov and L.~N.~Lipatov,
  {\it Deep inelastic e p scattering in perturbation theory},
  Sov.\ J.\ Nucl.\ Phys.\  {\bf 15}, 438 (1972)
  [Yad.\ Fiz.\  {\bf 15}, 781 (1972)].



\bibitem{Dokshitzer:1977sg} 
  Y.~L.~Dokshitzer,
  {\it Calculation of the Structure Functions for Deep Inelastic Scattering and e+ e- Annihilation by Perturbation Theory in Quantum Chromodynamics},
  Sov.\ Phys.\ JETP {\bf 46}, 641 (1977)
  [Zh.\ Eksp.\ Teor.\ Fiz.\  {\bf 73}, 1216 (1977)].

   \bibitem{Nazarenko}
S. Nazarenko, Wave Turbulence, Lecture Notes in Physics. Vol. 825, Springer, 2011.

\bibitem{Mehtar-Tani:2014yea} 
  Y.~Mehtar-Tani and K.~Tywoniuk,
  {\it Jet (de)coherence in Pb-Pb collisions at the LHC},
  [arXiv:1401.8293 [hep-ph]].

  



\bibitem{Arnold:2009ik} 
  P.~B.~Arnold, S.~Cantrell and W.~Xiao,
  {\it Stopping distance for high energy jets in weakly-coupled quark-gluon plasmas},
  Phys.\ Rev.\ D {\bf 81}, 045017 (2010)
  [arXiv:0912.3862 [hep-ph]].
 

\bibitem{Blaizot:2013vha}
  J.~P.~Blaizot, F.~Dominguez, E.~Iancu and Y.~Mehtar-Tani,
{\it Probabilistic picture for medium-induced jet evolution},
  JHEP {\bf 1406} (2014) 075
  [arXiv:1311.5823 [hep-ph]].
  
\bibitem{Fister:2014zxa} 
  L.~Fister and E.~Iancu,
{\it Medium-induced jet evolution: wave turbulence and energy loss},
  arXiv:1409.2010 [hep-ph].
 

\bibitem{Kurkela:2014tea} 
  A.~Kurkela and E.~Lu,
{\it Approach to Equilibrium in Weakly Coupled Non-Abelian Plasmas},
  Phys.\ Rev.\ Lett.\  {\bf 113}, no. 18, 182301 (2014)
  [arXiv:1405.6318 [hep-ph]].
  
  
  \bibitem{Ghiglieri:2015ala} 
  J.~Ghiglieri, G.~D.~Moore and D.~Teaney,
  JHEP {\bf 1603}, 095 (2016)
  doi:10.1007/JHEP03(2016)095
  [arXiv:1509.07773 [hep-ph]].
  
  
  \bibitem{Berges:2013eia} 
  J.~Berges, K.~Boguslavski, S.~Schlichting and R.~Venugopalan,
  Phys.\ Rev.\ D {\bf 89}, no. 7, 074011 (2014)
  doi:10.1103/PhysRevD.89.074011
  [arXiv:1303.5650 [hep-ph]].
  
  \bibitem{Tanji:2017suk} 
  N.~Tanji and R.~Venugopalan,
  Phys.\ Rev.\ D {\bf 95}, no. 9, 094009 (2017)
  doi:10.1103/PhysRevD.95.094009
  [arXiv:1703.01372 [hep-ph]].
  
  \bibitem{Tanji:2017xiw} 
  N.~Tanji and J.~Berges,
  Phys.\ Rev.\ D {\bf 97}, no. 3, 034013 (2018)
  doi:10.1103/PhysRevD.97.034013
  [arXiv:1711.03445 [hep-ph]].
 

%
\end{thebibliography}



\end{document}